\documentclass[8pt]{article}

\usepackage[english]{babel}

\usepackage{amsmath}
\usepackage{amsfonts}
\usepackage{graphicx}
\usepackage{parskip}
\usepackage{comment}
\usepackage{tcolorbox}
\usepackage{tabularx}

\let\oldabstract\abstract 
\let\oldendabstract\endabstract
\makeatletter
\renewenvironment{abstract}
{%
               {\list{}{\addtolength{\leftmargin}{-1em} 
                        \listparindent 1em%
                        \itemindent    \listparindent%
                        \rightmargin   \leftmargin%
                        \parsep        \z@ \@plus\p@}%
                \item\relax}%
               {\endlist}%
\oldabstract}
{\oldendabstract}
\makeatother

\usepackage{geometry}
\usepackage{xcolor}
\usepackage{hyperref}
\usepackage{cleveref}
\usepackage{subfig}
\usepackage{booktabs}
\usepackage{graphicx}
\usepackage{titlesec}
\usepackage{sectsty}
\sectionfont{\raggedright}
\titlespacing\section{1pt}{12pt plus 3pt minus 2pt}{1pt plus 2pt minus 2pt}
\titlespacing\subsection{1pt}{12pt plus 3pt minus 2pt}{1pt plus 2pt minus 2pt}
\titlespacing\subsubsection{1pt}{12pt plus 3pt minus 2pt}{1pt plus 2pt minus 2pt}

\definecolor{ForestGreen}{RGB}{34,139,34}
\definecolor{MidnightBlue}{RGB}{25, 25, 112}

\hypersetup{
    colorlinks=true,
    urlcolor=MidnightBlue,
    linkcolor=ForestGreen,
    citecolor=ForestGreen
}

\usepackage[
backend=biber,
style= authoryear,
]{biblatex}

\usepackage{multicol}
\bibliography{sample.bib}

 \newcommand\blfootnote[1]{%
  \begingroup
  \renewcommand\thefootnote{}\footnote{#1}%
  \addtocounter{footnote}{-1}%
  \endgroup
}

\newenvironment{Figure}
  {\par\medskip\noindent\minipage{\linewidth}}
  {\endminipage\par\medskip}

\usepackage{scalerel}
\usepackage{stackengine,wasysym}

\def\sym#1{\ifmmode^{#1}\else\(^{#1}\)\fi}

\title{The Compute Divide in Machine Learning: A Threat to Academic Contribution and Scrutiny?\vspace{-2ex}}

\author{\small 
  Tamay Besiroglu\footnotemark[2] \footnotemark[5] , Sage Andrus Bergerson\footnotemark[1] , Amelia Michael\footnotemark[2]  \footnotemark[3] , Lennart  Heim\footnotemark[5] \hspace{0.01em} \footnotemark[6], 
  Xueyun Luo\footnotemark[2] \footnotemark[4],  \hspace{0em} Neil Thompson\footnotemark[2]  \\ 
  \small{\footnotemark[1] New York University, \footnotemark[2] MIT FutureTech, \footnotemark[3] Brown University,}\\  \small{\footnotemark[4] Cornell University, \footnotemark[5] \hspace{0.1em}Epoch, \footnotemark[6] Centre for the Governance of AI}
}
\date{
}
\begin{document}
\maketitle
\begin{multicols}{2}
\begin{abstract}
There are pronounced differences in the extent to which industrial and academic AI labs use computing resources. We provide a data-driven survey of the role of the compute divide in shaping machine learning research. We show that a compute divide has coincided with a reduced representation of academic-only research teams in compute intensive research topics, especially foundation models. We argue that, academia will likely play a smaller role in advancing the associated techniques, providing critical evaluation and scrutiny, and in the diffusion of such models. Concurrent with this change in research focus, there is a noticeable shift in academic research towards embracing open source, pre-trained models developed within the industry. To address the challenges arising from this trend, especially reduced scrutiny of influential models, we recommend approaches aimed at thoughtfully expanding academic insights. Nationally-sponsored computing infrastructure coupled with open science initiatives could judiciously boost academic compute access, prioritizing research on interpretability, safety and security. Structured access programs and third-party auditing may also allow measured external evaluation of industry systems.

\end{abstract}

The crucial role of powerful computer hardware that is well-suited to the needs of machine learning has been recognized in the field for at least twenty years. In the context of neural-network-based machine learning, it has been discussed since at least the mid-2000s (see, in particular, \cite{uetz2009large, cirecsan2011handwritten, chellapilla2006high}). The ascent of deep learning has cemented this  recognition (\cite{lecun2015deep, schmidhuber2015deep}). Work on large self-supervised models, and particularly large language models (LLMs)—such as GPT-4 and PaLM—has underscored the salience of large-scale AI computing resources as a critical input for training capable deep learning systems. Self-supervised learning, in particular, has enabled more scalable training on massive unlabelled datasets, which has resulted in compute becoming one of the key bottlenecks in the development of state-of-the-art machine learning models (\cite{bommasani2021opportunities}).

\blfootnote{We are grateful to the participants at the July 2022 Compute Divide workshop hosted at the FutureTech Project at CSAIL MIT, to Juan Mateos-Garcia, Micah Musser, Konstantin Pilz, Phillip Isola, Jayson Lynch, Matthew Zaragoza-Watkins, Ben Garfinkel, Markus Anderljung, Micah Musser, Nur Ahmed, and Michael Totty. for helpful discussions. We thank Owen Dudney for his help in producing the NeurIPs compute usage dataset.}
\begin{Figure}
    \centering
    \includegraphics[width=0.95\linewidth]{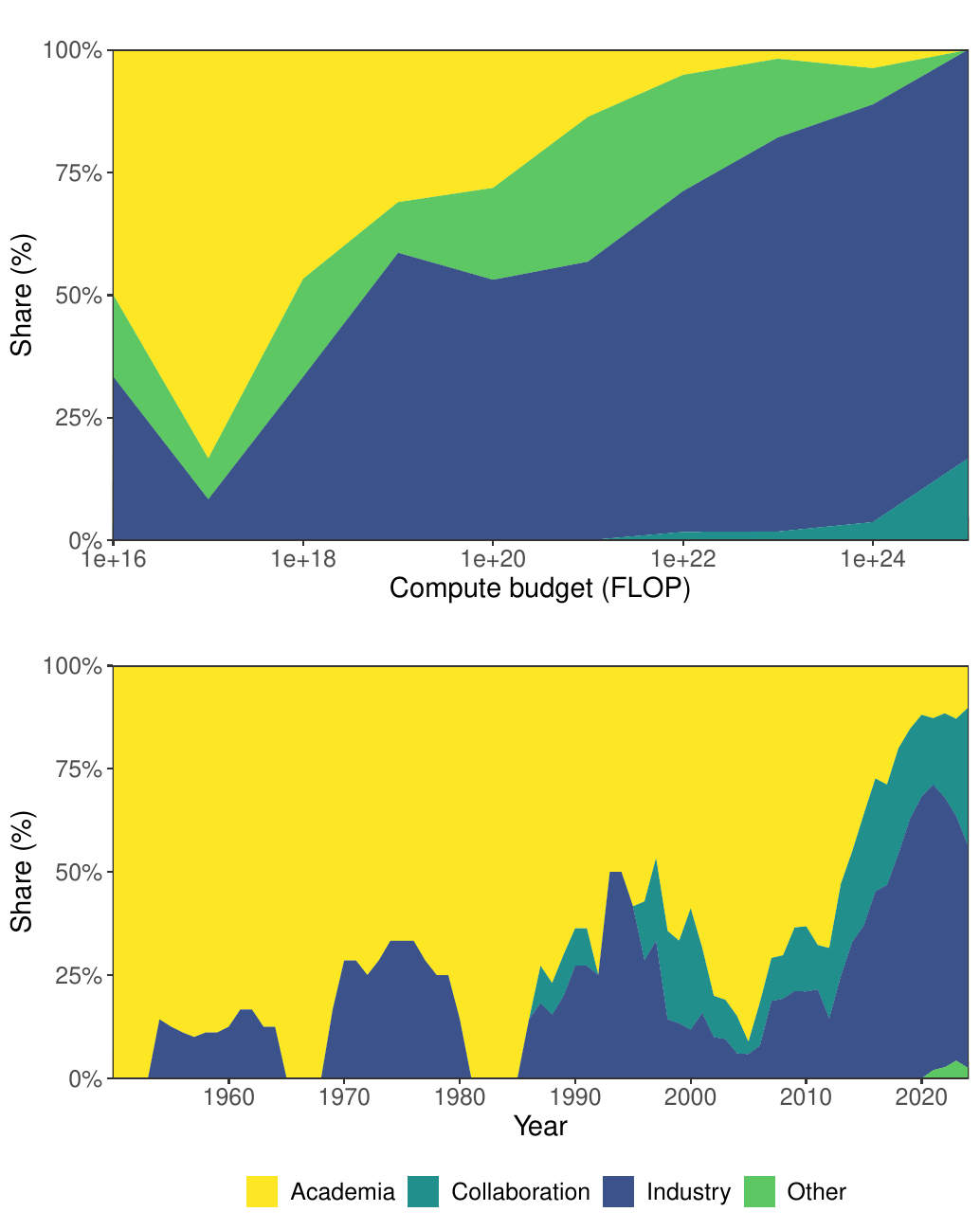}
    \label{fig:figure1}
    \captionof*{figure}{\footnotesize \textbf{Figure 1.} Share of notable machine learning models by institution type at various compute budgets (top).\footnotemark Machine learning models in the top 75th percentile of compute intensity relative to contemporaneous models by institution type over time (bottom). Based on a dataset of 650 machine learning models across a broad range of domains by \cite{sevilla2022compute}.}
\end{Figure}\footnotetext{“Notable” is defined as results that satisfy at least one notability criterion (90th percentile of citations among AI publications from the same period, historical importance or a state-of-the-art advance), see \hyperref[AppendixA]{Methods}.}
The compute required to train machine learning models, measured in floating point operations (FLOP), has doubled every six months on average since the early 2010s (\cite{sevilla2022compute}). By contrast, Moore’s law is typically associated with a two-year doubling time. Since late 2015, a new class of unusually large models that use 10- to 100-fold more compute than their typical contemporaries has become a common fixture in AI. Estimates indicate that the amount of compute needed to train large-scale models has increased between hundreds of thousands and tens of million-fold in the past decade (\cite{sevilla2022compute, ganguli2022predictability}). With this growth in compute, the training of widely used models has become increasingly technically demanding and expensive, with costs reaching excess of tens of millions of dollars. Meanwhile, the cost of workhorse language models, such as the medium to large size BERT, can already be too costly for projects within academia (\cite{sharir2020cost, izsak2021train}).
 
The increasing amount of compute also brings greater technical complexity in its deployment, requiring expertise in parallel and distributed computing, specialist programming languages, and software development and optimization. Hardware constraints, including memory limitations and communication bandwidths on accelerators, require practitioners to adopt various distributed computing techniques, notably model parallelism, which are often difficult to design and implement (\cite{huang2019gpipe, shoeybi2019megatron}). Hardware failures in training often require manual remediation from dedicated engineers (\cite{zhang2022opt}), an issue that is likely to be especially challenging for academic machine learning groups that might lack engineers who specialize in the hardware used in research experiments.

Despite discussions on the divide in research contributions due to compute access (\cite{ahmed2020democratization, ahmed2023growing}), a comprehensive account of its implications for machine learning and AI governance is lacking. In light of this, we provide a data-driven investigation into the way compute divides machine learning research and offer a preliminary analysis of the strategic implications. We ask: How does a widening compute divide affect the machine learning research ecosystem? What are the consequences for the ability of the scientific community to access, evaluate, study, and scrutinize machine learning models and other artifacts?

\section{Compute divides machine learning research}\label{sec:sec2}

The dominant position once held by academia in the development of significant machine learning models appears to have declined recently. An analysis of 650 notable machine learning models indicates that the proportion of large-scale machine learning models (characterized as those in the top quartile of computational usage compared to their counterparts) created by academic laboratories has decreased dramatically. In the early 2010s, around 65\% of these models originated from academic labs. However, this figure fell to approximately 10\% during the initial years of the 2020s (as shown in Figure \ref{fig:figure1}). Moreover, since 2017, industry-only research teams have dominated the training of large-scale machine learning models, reaching around 81\% in 2022. 

We find that the relative dominance of industry is especially pronounced in the case of large self-supervised models, such as language or generative image models, where non-industry models are rare and generally much smaller than those produced by industry. The largest self-supervised model trained by an academically led research team is ProtT5-XXL (\cite{elnaggar2020prottrans}), which, using supercomputers from the Oak Ridge National Laboratory, still used less than 1\% of the compute used to train the largest machine learning model developed by industry. Another significant shift in the contributions from various institutions is the rise in collaborative efforts, which have accounted for about 20\% of large-scale machine learning models since 2012. This is a notable increase from the single-digit percentages observed in the early 2000s. Such collaborations often serve as a bridge for knowledge exchange or as a conduit for academic researchers transitioning to industry roles. As a result, this trend could be viewed as a precursor-trend that has enabled the current relative dominance of industry.

Of the large models trained outside of industry, a notable fraction is produced by Chinese research organizations. This suggests that the compute divide may be less pronounced in Chinese AI development. Indeed, Chinese government-sponsored labs, such as the BAAI, provide an unrivaled level of support in terms of funding and compute to facilitate the training of large machine learning models (\cite{ding2022recent}).
\vspace{-0.5cm}
\begin{Figure}
    \centering
    \includegraphics[width=0.7\textwidth]{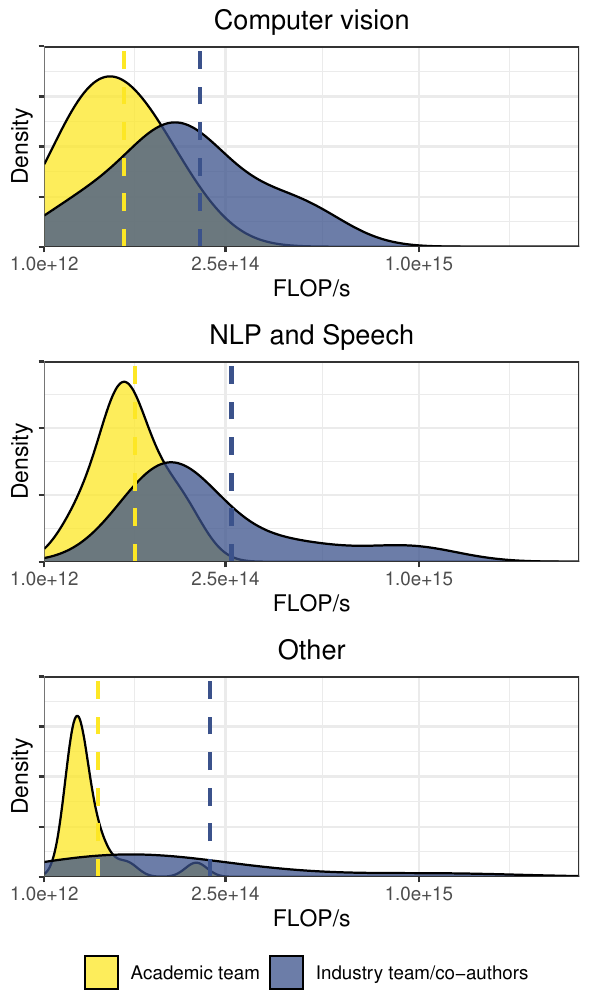} 
    \label{fig:example}%
    \vspace{-0.25cm}
    \label{fig:figure 2}
    \captionof*{figure}{\footnotesize \centering \textbf{Figure 2.} Distribution of peak theoretical performance (FLOP/s) of hardware used in 2,333 NeurIPS 2021 papers. Dashed lines show mean peak theoretical performance of hardware set-ups reported in papers by academic research teams and research teams that involve industry authors. ``Other" refers to topics other than NLP and speech, and computer vision (e.g. reinforcement learning, methodological papers, graph neural networks, and papers covering multiple subfields).}
\end{Figure}

Industry-affiliated research groups report using hardware five times more powerful than non-industry groups in NeurIPS publications. Around 30\% of industry teams also have better infrastructure than the best academic setups. While some industry teams have similar resources to academics, highly compute-intensive projects are predominantly done by industry. Overall, the current picture-of labs producing large-scale models and those using large amounts of computing resources-is consistent with the pronounced differences between industry and non-industry labs in their ability to fund or deploy high-performance computing resources.

\begin{itemize}
    \item \textbf{Funding.} A key reason for the compute divide between academia and industry is the limited funding available for researchers in academia. The high costs of purchasing or renting many AI accelerators (such as GPUs) may encumber academic researchers from acquiring the resources needed for compute intensive machine learning research.

    \item \textbf{Engineering expertise.} The setup and management of large clusters of hardware and AI accelerators needed for compute-intensive research requires extensive expertise in parallel and distributed computing (\cite{huang2019gpipe, shoeybi2019megatron}). Orchestrating these large clusters is its own discipline and requires dedicated staff. This type of work is not commonly found in academia, as PhD research in machine learning is often done by individuals or small teams.
    
    \item \textbf{Access to computing clusters.} Industry players, especially large tech firms, are often supplied or co-located with access to data centers and given these services for free or at reduced prices. It is, for instance, notable that many of the leading industry AI research groups are also leading cloud providers, such as Amazon Web Services, Google Cloud, and Microsoft Azure (\cite{gartner2021iaas, Belfield}), or at least have partnerships with cloud providers. The availability, ease of access, and on-site talent for leveraging these computational resources make it easier for industry researchers to use these resources for their research and development efforts.

    \item \textbf{Academic research is more diverse.} Academic research is often more diverse and may not always require large amounts of compute resources. Industry, in contrast, typically has a more single-minded focus on commercializing their research and may be more interested in computationally intensive projects (\cite{klinger2020narrowing}). While minor improvements in an already extensively explored domain are of interest to industry, research in well-explored domains may offer fewer of the novel scientific insights sought after in academia.
\end{itemize}
\section{Implications for academic contributions}

The growing divide in compute resources significantly impacts the distribution of research topics between industrial and academic institutions in ML domains. In areas where compute usage and the need for high-performance computing expertise rise, industry is taking the lead. In response, academia is increasingly focusing on lower-compute intensity research, and research involving open source pre-trained models developed by industry.

\subsection{Impact on research landscapes}
Using publication data from \href{https://openalex.org/}{OpenAlex}, we find that an increase in the number of compute-related terms (i.e. ``GPU", ``large-scale", ``distributed computing", etc.) in ML sub-fields is associated with a decline in the proportion of research produced by academic-only research teams.
\begin{Figure}
    \centering
    \includegraphics[width=1\textwidth]{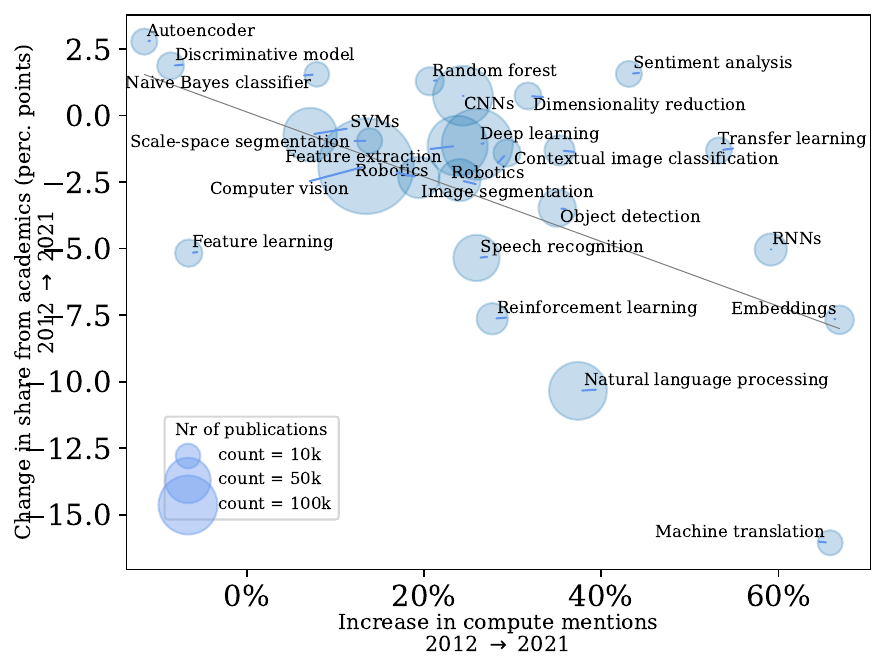} 
    \label{fig:example}%
    \label{fig:figure3}
    \vspace{-0.8cm}
    \captionof*{figure}{\footnotesize \centering \textbf{Figure 3.} As the average number of compute-related terms in abstracts for papers by topic increases, the share of research publications on those topics from academic-only teams decreases (slope: -0.122, p-value: $<$ 1\%). Plot shows top 25 ML topics in the OpenAlex corpus.}
\end{Figure}
In the top 25 ML sub-fields ($\sim$1.2M publications), we observe that in the 2012-2021 period, a 10\% rise in compute-related keywords in abstracts corresponds to a 1.73\% decrease in the proportion of academic-only publications in a topic. This effect is particularly noticeable in language-related ML fields: a 10\% increase in compute-related keywords in abstracts is linked to a reduction in academic-only team representation in natural language processing by 2.43\%, in language modeling by 2.76\%, and in machine translation by 2.18\%.

The data also indicates that an overall increase in compute usage can be differentially attributed to industrial AI labs. While compute-related key words in ML abstracts grew on average 17.8\% in the 2012-2021 period in academic abstracts, this number was 33.0\% in industry abstracts. This differential increase was especially pronounced in domains such as robotics, transfer learning, reinforcement learning, and natural language processing (Figure 4).

\begin{Figure}
    \centering
    \includegraphics[width=1\textwidth]{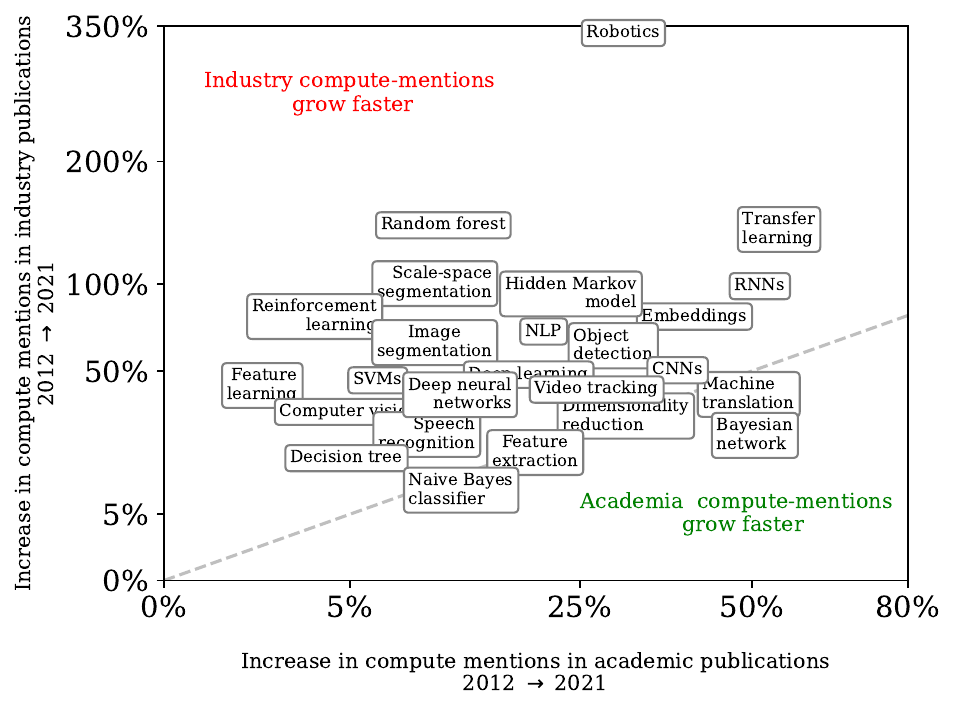} 
    \label{fig:example}%
    \label{fig:figure4}
    \vspace{-0.8cm}
    \captionof*{figure}{\footnotesize \centering \textbf{Figure 4.} Change in compute mentions for the top 25 sub-fields of ML by publications that saw an increase in compute-mentions over the 2012-2021 period according to \href{https://openalex.org/}{OpenAlex} data.}
\end{Figure}

\subsection{Sidelining from research on foundation models}

The massive training runs of large models require extensive engineering expertise, and practitioners must often adopt various types of distributed computing techniques, notably model parallelism, that are hard to design and implement (\cite{huang2019gpipe, shoeybi2019megatron}). The tooling for implementing effective large-scale training, such as model parallelism, tensor parallelism, or effective reductions in memory consumption, is often nascent at a  large-scale and relies on custom frameworks (\cite{athlur2022varuna}).

Current data strongly suggests that the driving force behind research into large self-supervised models is predominantly the industrial sector, as opposed to academic research institutions. This is exemplified by the significant gap in the production of these models. Of the 23 models with over a billion parameters released in 2022, industrial research labs were solely responsible for producing 16 of these models, while academic institutions could only claim exclusive credit for one (Figure 5).

Academic research often aims to advance basic knowledge, while industry research tends to focus on developing capabilities for commercial products (\cite{muller2014}, \cite{bodas2017motivations}). Although industry faces some pressure to align with academic interests for recruiting talent (\cite{ahmed2022scientific}), they may attract academics through other means like salaries, computing resources, and access to proprietary datasets.

\subsection{Dominance of industry open sourced models}

Open source programming languages (e.g. Python), frameworks (e.g. PyTorch, TensorFlow), software, and datasets are widely used in computer science research (\cite{barba2022defining, 9153295}), and ML in particular (\cite{langenkamp2022open}). Aided by platforms such as \href{https://huggingface.co/}{HuggingFace}, open source ML models are becoming another important tool in ML research. As developing large self-supervised models from scratch is infeasible for most academic labs due to the relative lack of compute access, we expect to see an increased emphasis on researching open sourced pre-trained models.

In addition to demand for pre-trained models from non-industry researchers, incentives to regularly release open source ML model weights and code-bases exist in industry, suggesting this trend will continue. Open-sourcing technology can provide benefits to the developers, such as reducing the costs of testing, debugging, and generating improvements (\cite{lerner2002some, henkel2004open, aagerfalk2008outsourcing}), lowering the recruiting and on-boarding costs of new hires by promoting adoption of one’s technology stack (\cite{lerner2002some, marlow2013activity}), as well as allowing for the sale of complementary products (\cite{dahlander2006man, watson2008business}).
\begin{Figure}
    \centering
    \includegraphics[width=1\textwidth]{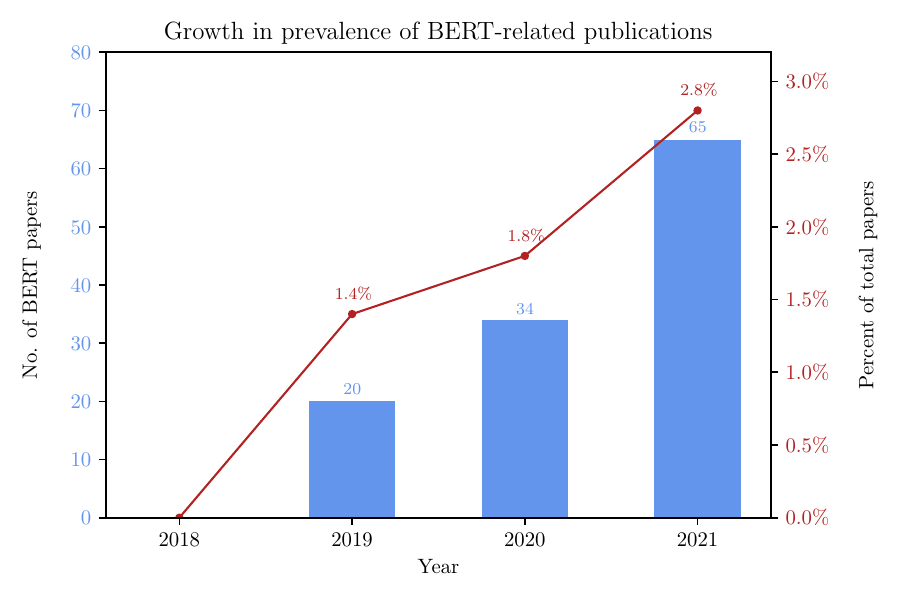}
    \label{fig:example}%
    \vspace{-2.5em}
    \captionof*{figure}{\footnotesize \centering \textbf{Figure 5.} Total count and proportion of papers at NeurIPS conferences that studied BERT after its 2018 release. All papers from NeurIPS 2018 - 2022 were scraped for mentions of BERT, RoBERTa, or DistillBERT. Those with at least one mention were isolated and analyzed for whether or not they sufficiently used or studied a BERT model or technique in their research.}
\end{Figure}
Open source ML models like BERT offer non-industry researchers-tested, large-scale models with lower computational costs and engineering overhead. A notable example is Google’s BERT (\cite{devlin2018bert}), which has become an ubiquitous baseline in NLP experiments and has given rise to ‘BERTology’, the study of the nature and performance of the model (\cite{rogers2020primer}). After BERT was released in 2018, the number of research papers studying BERT (and models in the BERT-family) at NeurIPS more than tripled from 2019 to 2021 (Figure 6).  By 2021, around 3\% of all NeurIPS publications (a total of 65 papers), investigated BERT models or conducted experiments on BERT.

In their analysis of interviews of open source ML software users, \cite{langenkamp2022open} highlight that open-sourcing in ML has facilitated the creation of standards, facilitating interoperability and fostering a growing user base. Open source ML models, such as BERT, may then form the basis of a set of standards for architectures, input/output representations, and other settings in a way that preempts the development of competing standards.

Corporate AI labs may aim to open source ML models to invite external researchers to test, evaluate and iterate on existing techniques. While the publication of code lowers the cost of reproduction and iteration, (\cite{langenkamp2022open}), while releasing model weights lowers these even more drastically. Since the availability of tools and models can shape researchers' research priorities, the release of models could promote research on topics of interest. For instance, \cite{zhang2022opt} noted that release of the OPT suite of models could advance progress on their ability to improve robustness and mitigate issues such as bias and toxicity through open collaboration with external researchers.

\subsection{Policy implications and recommendations}

The foregoing section highlights several policy-relevant implications of the compute divide, including reduced representation in compute intensive subfields, the training and evaluation of large-scale models, and industry dominance over open source artifacts. In what follows, we provide preliminary recommendations for promoting and supporting research that we may expected to be under-provided by industry labs.

\subsubsection*{Responsible compute provision}

There is a growing awareness of the resource gap in AI research within academia, prompting efforts to provide computing resources. Recent initiatives, such as the US National AI Research Resource (NAIRR; \cite{naarrtf2023strengthening}) and the UK's AI Research Resource (\cite{compute_review}) aim to, in part, mitigate the effects of limited compute access for academic institutions on national competitiveness and talent development. Nationally sponsored computing infrastructure can potentially achieve significant economies of scale in the procurement and maintenance of large-scale infrastructure (\cite{Anderljung2022}).

While these resources could help address the divide, it is crucial to pair increased compute allocation with a focus on risk reduction and equitable benefit distribution (\cite{Anderljung2022}), so that it expands academic insights judiciously, empowering academia to guide progress responsibly. We thus emphasize the importance of compute provision schemes aimed at supporting research that may reasonably be expected to be under-provided by industry labs, such as: 
\begin{enumerate}
\item Improve scrutiny, comprehension, and interpretability of large machine learning models;
\item Concentrate on enhancing AI safety, security, and robustness to address potential risks and vulnerabilities in high-compute models;
\item Create open source models to address challenges not prioritized by industry due to monetization barriers or lack of immediate financial returns (e.g. models for neglected tropical diseases, climate change, public health, or under-resourced languages); and
\item Encourage exploration of high-uncertainty/high-reward research, especially that which blends diverse approaches from traditionally academic-dominated fields, to foster breakthroughs and new paradigms in AI.
\end{enumerate}

\subsubsection*{Open science initiatives}

Promoting open science initiatives that prioritize the sharing of pre-trained models can significantly bridge the compute divide, particularly for projects focusing on public benefits or addressing issues that are likely to be underserved by the industry.\footnote{An example of such efforts is \cite{bigscience_bloom_lm}, supported in part by GENCI, a public organization promoting intensive computing and AI use for French research communities.} 

Support should prioritize research that directly tackles challenges not prioritized by industry due to monetization barriers or lack of immediate financial returns (e.g. such as experiments aimed towards understanding risks from AI, models for neglected tropical diseases, climate change, public health, or under-resourced languages) as well as other applications of AI that serve the common good (such as those discussed in \cite{pizzi2020ai, hager2019artificial}).

\section{Scrutiny and diffusion of ML models}

A compute divide between academia and industry impacts the diffusion and distribution of machine learning systems, and the degree to which they undergo scrutiny, critical evaluation, and testing.

\subsection{Diffusion of machine learning models}
\label{sec:4.1}

The compute divide may constrain the diffusion of machine learning models, as prominent AI labs like Google Research and OpenAI may withhold their state-of-the-art models (\cite{Talat_etal_2022, Wiggers_2022}). Firms might maximize commercial value of technology by minimizing "knowledge spillovers" (\cite{cassiman2002r, alexy2013cui}), thus discouraging the unrestricted release of research artifacts such as models, code bases, and data. 

On the other hand, there are a host of reasons corporate labs might publish technical work, because this can:
\begin{itemize}
\item signal tacit knowledge, enhance technical reputation for fundraising, recruiting, and other purposes (\cite{hicks1995published, ahmed2022scientific});
\item defensively block consolidation or patenting by competitors (\cite{barrett2002defensive});
\item support marketing activities and market positioning through demonstrating technical competency to investors or consumers (\cite{azoulay2002pharmaceutical, polidoro2012getting});
\item exploit second-mover advantages by subsidizing first-movers (\cite{dai2022strategic}); and
\item avoid consolidation by a rival (\cite{aguelakakis2019collaborate}).
\end{itemize}

\begin{Figure}
    \centering
    \includegraphics[width=1\textwidth]{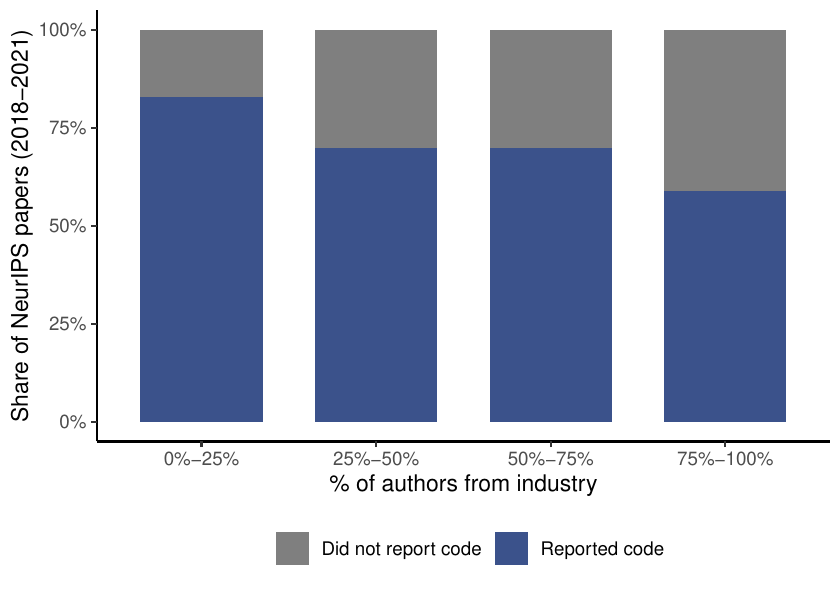}
    \label{fig:example}%
    \vspace{-2.5em}
    \captionof*{figure}{\footnotesize \centering \textbf{Figure 6.} A higher fraction of industry-affiliated authors on NeurIPS papers is associated with less frequent publication of the relevant code base. Linear regressions are fit on a random sample of 519 NeurIPS papers (see \hyperref[AppendixE]{Methods}).}
\end{Figure}

Nevertheless, the data on code-sharing suggests that industry tends to be more tight-lipped about its work than academic teams. We find a negative relationship between industry authorship and the rate of code release. In particular, for a sample of 565 NeurIPS publications, we find that industry-only teams are about 32\% less likely to release their code than a paper with academic-only authorship. This negative relation holds across all years of publication in our data set. This finding is consistent with our predictions of higher transparency in academia, given the theoretical differences in incentive structures between academia and industry.

Transparency, verifiability, replicability, and openness are deeply ingrained principles in norms and practices of academic science broadly (\cite{munafo2017manifesto, resnik2005ethics}), and in AI in particular (\cite{whittlestonetension}), fostering a broader diffusion of research. Academic researchers largely prioritize novelty, prestige, research prizes, and academic success over profit, making certain industry-related disincentives against knowledge diffusion less applicable. However, it is essential to consider the impact of cultural norms and reputational concerns on these incentives. Academic AI labs may hesitate to share models that could potentially be misused or cause harm due to concerns about negative consequences with funders, peers and the public. Furthermore, first-mover advantages, in which early publishers receive disproportionately higher citations (\cite{newman2009first}), can occasionally clash with the prompt dissemination of knowledge.

Our analysis suggests that industry labs face notable disincentives for diffusion, such as protecting intellectual property and mitigating risks misuse and unintentional harm and the associated reputational risks. Although these disincentives are not entirely unique to the industry, the relatively strong track record of academic machine learning research for openness, as supported by our data on code-sharing, suggests that a compute-divide-induced dominance of industry labs will reduce the extent of diffusion of certain machine learning artifacts compared to the counterfactual if compute was disproportionately in the hands of academic labs.

\subsection{Scrutiny of machine learning models}

Scrutiny of machine learning models and associated techniques—through testing, validation, and critical evaluation—is crucial to understanding the limitations and risks of AI systems. Evaluating how an industry dominance of compute intensive AI research affects the level of scrutiny is key to understanding the trend's governance implications.

The scrutiny provided by industry and academia can have different and complementary emphases. Because industry labs focus predominantly on AI systems for deployment in products or applications, we expect industry scrutiny to be more holistic—focusing on how such a system behaves, at scale, when embedded within some human-computer interactive setting. By contrast, such `embedded' contexts in large-scale deployments are accessible in academic research. 

Industry labs can have substantial incentives to provide such scrutiny; there can, as there could be a large value in understanding the limitations of AI systems if that enables the building of more highly economically useful models (see, e.g., \cite{brundage2021lessons, lowe2022aligning} for a discussion of how safety-promoting research aligns with improving capabilities). Yet, it is unclear whether the incentives industry faces are adequate. When compared to other industries, AI development might, at least in the near future, involve weak incentives for providing internal scrutiny. Because consumers lack understanding of AI-associated risks—which can be unusually challenging to understand (\cite{askell2019role})—traditional market incentives for delivering safe products are weaker. A comparable lack of understanding among regulators makes safety mandates difficult to craft and enforce. Risks may be diffuse, like reduced online trust, so firms could fail to internalize costs of rushed deployment. And first-mover advantage may reward rapid development over time-consuming safety-evaluations evaluation (\cite{askell2019role}).

Given the empirical observations described in Section \ref{sec:4.1}, we may expect the lopsided role of industry in the production of compute intensive machine learning systems to limit the diffusion of relevant artifacts. As noted elsewhere (such as by \cite{liang2022community-norms, whittlestonetension, solaiman2019release}), more limited diffusion may reduce the extent of external scrutiny. More broadly, the reduced accessibility of the relevant artifacts will hamper the research community's academic and industry and government bodies' ability to test, validate, and develop a safety-promoting understanding of models and associated techniques.

Research that scrutinizes compute intensive machine learning models and the associated techniques is plausibly well-suited to academia. First, academic research can be conducted at arms-length from organizations deeply invested in the technologies, which can be crucial for critiquing approaches pursued by industry labs. For example, in their work, \cite{bender2021dangers}, evaluates the shortcomings and risks of large language models. It is reported that this publication resulted in one of the Google-employed authors having their employment terminated (\href{https://www.wired.com/story/behind-paper-led-google-researchers-firing/}{Wired, 2020}). Examples such as this suggest that the independence of academic researchers, and perhaps particularly tenured researchers, might be critical in conducting work aimed at providing scrutiny. Second, academic research labs are more likely to publish work that uncovers the risks and limitations of their models. While industry labs might be disinclined to publish such work, as doing so could disclose sensitive intellectual property or it might harm their reputation and market positioning, academia tends to be more geared towards public research outputs. While this may not always promote safety (e.g. publishing prompt-injection techniques), public research that advances a deeper understanding of the relevant issues is arguably ultimately more likely to result in desirable outcomes than ``security through obscurity''.

\subsection{Policy implications and recommendations to address scrutiny}

As this work emphasizes, academic researchers can possess the necessary expertise, perspective and independence to scrutinize, test and evaluate compute intensive ML models effectively. As AI systems continue to grow in capability and ubiquity, ensuring that academia can investigate how these systems could potentially fail, inflict harm, or generate unintended consequences becomes increasingly vital. Insofar as academia is likely to fall short of being able to contribute these goals, other mechanisms will need to be considered.

\subsubsection*{Privileged structured access}

Structured access, as presented in \cite{shevlane2022structured}, is an emerging paradigm for safely deploying artificial intelligence systems. Rather than openly distributing AI software, developers facilitate a controlled interaction with the AI system’s capabilities, using technical and other methods to limit how the software can be used, modified, and reproduced. There have been various examples of this, such as GPT-3 (\cite{brown2020language}) and OPT-175B (\cite{zhang2022opt}), both of which had structured access programs for researchers in academia and those affiliated with governmental or civil society institutions.

By granting researchers in academia and other non-industry labs structured access, industry models can be evaluated more widely and studied independently by AI researchers, as well as by social scientists, ethicists, and other non-technical experts. Independent evaluation can reveal potential biases, flaws, or unintended consequences in the AI systems that might have been missed by the developers, thereby helping to improve their quality and safety. Importantly, structured access can be consistent with industry aims of preserving ownership over intellectual property, though this might require carefully designed protocols.

Standardized protocols for privileged structured access programs are currently lacking, and many questions on how such programs should be designed are still unresolved, such as:
\begin{itemize}
    \item What technical and other methods for limiting how the model can be used, modified, or reproduced strike the right balance between enabling thorough research and preserving intellectual property?
    \item How can we ensure that a diverse range of researchers and institutions have equitable access to structured access programs, avoiding the risk of further entrenching existing inequalities in the AI research community?
    \item What measures should be taken to prevent misuse of AI systems accessed through structured access programs, and what responsibilities do stakeholders have in monitoring and addressing such misuse?
    \item How can structured access programs be adapted and updated to respond to the rapidly evolving AI landscape, ensuring they remain relevant and effective as technology advances?
\end{itemize}
We recommend academia, industry, and other stakeholders develop standardized protocols for granting structured access and ensuring transparency and consistency across different AI systems -- e.g., in the US the NAIRR might be particularly well positioned for setting this standardized protocol. Establishing a common framework for sharing access to high-compute models will create an environment that fosters independent scrutiny and evaluation of AI systems while maintaining a balance between intellectual property rights and the need for public accountability.

\subsubsection*{Third-party auditing}
As we have argued, academia will likely play a smaller role in providing critical evaluation and scrutiny to the machine learning artifacts produced by industry, especially large compute intensive models. In light of this, it is important to create other procedures to ensure that such artifacts still receive necessary scrutiny as systems become more widely deployed.

One such alternative is third-party auditing, a process where an external, independent auditor evaluates an AI developer's claims about their system's safety, security, privacy, and fairness (\cite{brundage2020toward}). Third-party auditing can help verify the accuracy of the developer's claims and promote consistency and accountability in the AI industry.

We recommend that stakeholders, including AI developers, policymakers, and civil society organizations, create a task force to research options for conducting and funding third-party auditing of AI systems. A task force focused on this issue could explore appropriate initial applications to audit, devise approaches for handling sensitive intellectual property, and balance the need for standardization with the need for flexibility as AI technology evolves. Collaborative research into this domain seems especially promising given that the same auditing process could be used across labs and countries.

Third-party auditing can complement the role of academia in providing critical evaluation and scrutiny of AI systems. By ensuring that independent auditors have the necessary expertise, resources, and access to evaluate AI systems effectively, we can create a robust ecosystem that fosters transparency, trust, and accountability in the development and deployment of AI technologies. Moreover, third-party auditing could serve as a bridge between AI developers and academia, allowing academic researchers to engage with industry models and contribute their expertise without being directly involved in the development process.

\end{multicols}
\begin{multicols}{2}

\section{Conclusion}
As the AI landscape undergoes rapid transformations, pressing questions emerge: Who is shaping the technology behind important machine learning models? What organizations possess the ability to test, evaluate, and scrutinize research artifacts? How are the directions of machine learning research determined? In our view, two salient interconnected trends must be considered: the diminishing role of academic research in compute intensive domains and the dwindling scrutiny faced by powerful, influential models.

Addressing these challenges calls for deliberate interventions that aim to expand academic insights judiciously, empowering academia to guide progress responsibly. We explore solutions like responsible compute provision, open science, structured access, and third-party auditing. While not exhaustive, we hope this inspires further dialogue and innovation towards a responsible, transparent, and collaborative AI research community prioritizing critical evaluation. 

\section*{Methods}
\subsection*{Data on machine learning models and compute estimates}
\label{AppendixA}

We use the data from \cite{sevilla2022compute}, which includes over 650 machine learning models presented in academic publications and relevant gray literature that have an explicit learning component, showcase experimental results,  and meet at least one notability criterion (having at least 1000 citations, historical importance, a SotA performance advance, deployment in a notable context, or wide use).

An issue with these notability criteria in our context is that models deployed in a notable context or have wide use are disproportionately models developed by industry labs, as industry labs have more immediate access to a large client-base. Indeed, many of the models that meet only these conditions are industry-developed (e.g. Stable Diffusion, DALL-E Codex, and AlphaFold2). Therefore, we only consider models that are SotA improvements, of historical relevance according to \cite{sevilla2022compute}, or are highly cited (defined as receiving as many citations as the top-90\% most cited machine learning publications from that 5-year period.

\subsection*{Compute divide in NeurIPS proceedings}
\label{AppendixB}

To investigate how compute usage differs between academic and industrial labs at NeurIPS 2021, we randomly sampled 519 publications and for each, annotated the topic of the publication (e.g. NLP, vision or other), whether these publications had industry co-authors, whether the publication had a co-author from a Chinese AI lab, and the total amount of compute that was used to train the model. In total, we were able to infer the compute requirements for 109 models. The methodology for calculating the total amount of compute used to train the model we use is that described by \cite{sevilla2022compute}.

\subsection*{Thematic differences in engineering topics in NLP research}
\label{AppendixC}

We contrast the prevalence of topics in research across top academic and industry natural language processing publications. We see a stark difference in focus between top academic publications and high-profile industry publications. In particular, the latter frequently features engineering-related topics (such as training parallelism, memory efficiency, and training stability), while these are mostly absent from top academic publications. 

\subsection*{The rise of BERT-ology}
\label{AppendixD}

 In this analysis, papers which, for example, adapted or augmented BERT architecture or techniques to make a new model (\cite{sung2021training, ma2021luna}), investigated or improved a specific aspect of the BERT architecture or techniques (\cite{michel2019sixteen, mukherjee2020uncertainty}) or studied changes in BERT performances due to changes in input (\cite{hase2021out, xie2020unsupervised}) were classified as studying BERT. On the other hand, papers which, for example, tested performance of new data types in a variety of models including BERT (Rouhani et al., 2020), compared performance of an original model (not BERT-based) to a BERT-based model’s performance (\cite{lewis2020pre, akula2021robust}), or applied concepts from BERT to other scientific domains (\cite{reddy2021can}) were classified as not studying BERT.

\subsection*{Changes in research agendas}
\label{AppendixE}

We use \href{https://openalex.org/}{OpenAlex data} to study how academic and industry research teams are represented across research topics in machine learning. The approach we follow is that of 
\cite{Garcia2022}. In addition to the topics they include, we add additional topics that are labelled in the OpenAlex dataset. To generate a list of compute-related words, we extracted the abstracts of papers in the dataset from \cite{sevilla2022compute} that present models using at least $1\mathrm{e}19$ FLOPs and generated a wordcloud of the most commonly used words in those abstracts. From those, we select words that we expect to be associated with large-scale compute deployments.
\begin{tcolorbox}
        \begin{center}
            Compute-related terms\\  
        \end{center}
        \footnotesize  Scale, Parameters, GPT, Size, Billion, Parameter, Scaling,
        Computational, Compute, Latency, Memory, GPU, Graphics Processing Unit, Parallelism,
        Distributed Computing, Computation, Thousands, Gpus, Hardware,
        Million, Scaled, Days, Expensive, TPU, Costs, Gshard,
        Trillion, 175b, Billions, FLOPS, Infrastructure, Deeper, Engineering, Enormous, Massive, Months,
        Large-scale, Footprint, Budget, Megatron, 530B, CO2.
\end{tcolorbox}
In our code base, we have robustness checks that validate the general picture is robust to justifiable changes in which terms we select (i.e. that there is a strong negative association between compute terms in relevant abstracts, particularly in NLP, and the representation of academic teams).

\subsection*{Differences in code-release practices at NeurIPS proceedings}
\label{AppendixF}

To analyze the relative rates of diffusion between academia and industry, we analyze a random sample of NeurIPS papers presenting novel machine learning models. Our goal is to determine the effect of industry authorship on the likelihood of model code being published. Out of 565 papers in our sample, 94 present novel machine learning models. For each of these, we record the number of authors from industry and academia and whether the model’s code was made available (e.g., through a GitHub repository).
\begin{center}
\begin{center}
\begin{tabular}{lcccr}
\toprule
Variable & Estimate & Std. Error & Pr(\(\geq|z|\)) & \\ 
\midrule
Frac. industry & -1.52* & 0.719 & 0.035\\
No. of authors & -0.00253 & 0.106 & 0.981\\
2018 & 0.952 & 0.790 & 0.228\\
2019 & 1.30 & 0.674 & 0.054\\
2020 & 1.89* & 0.767 & 0.014 \\
2021 & 2.13** & 0.720 & 0.0031\\ 
\bottomrule
\end{tabular}
\textbf{Table 1.} Logistic regression results: effects of industry authorship and year on code publication. *, ** denote \(p<0.05\) and \(p<0.01\) respectively.
\end{center}
\end{center}

We find that the average proportion of industry authorship is 27\%, and the rate of code publication is 73\% (69 out of 94 papers). Using logistic regression, we find a significant negative relationship between the fraction of authors from industry and the likelihood of code release. Specifically, for a paper with 100\% industry authorship, the odds of releasing code are about 21.9\% of the odds for a paper with 0\% industry authorship (or 100\% academic authorship), assuming all other variables are held constant.

An increase from 0\% to 100\% industry authorship leads to
a reduction in the odds of reporting code of $1-\exp(-1.52) \approx 78\%$. Given that 81.40\% of all-academic teams report their code, this would amount to an absolute reduction in the probability of reporting by 32.45\%.

We also find that the odds of a paper releasing code increase over time. However, only the coefficients for the years 2020 and 2021 were statistically significant with an alpha of 0.05. These findings are consistent with our predictions of higher transparency in academia, given the theoretical differences in incentive structures between academia and industry. 

\end{multicols}

\printbibliography

\end{document}